\documentclass[preprint,prl,showpacs,amssymb,floatfix]{revtex4}
\newcommand{\be}{\begin{equation}}
\newcommand{\ee}{\end{equation}}
\newcommand{\beq}{\begin{eqnarray}}
\newcommand{\eeq}{\end{eqnarray}}
\usepackage{bm}
\usepackage{graphicx}%
\usepackage{epsf}
\usepackage{epsfig}

\usepackage{amsmath}
\usepackage{amsfonts,amssymb}

\newcommand{\Aslash}{\ensuremath \raisebox{0.025cm}{\slash}\hspace{-0.25cm} A}
\newcommand{\dslash}{\not{\hbox{\kern-2pt $\partial$}}}

\begin{document}
\title{Baryon currents in QCD with compact dimensions}
\author{B.~Lucini$^a$, A.~Patella$^{b,c}$ and C.~Pica$^{d}$}
\affiliation{$^a$ Physics Department, Swansea University, Singleton Park, Swansea SA2 8PP, UK\\ 
{$^b$ Scuola Normale Superiore, Piazza dei Cavalieri 27, 56126 Pisa, Italy}\\
{$^c$ INFN Pisa, Largo B. Pontecorvo 3 Ed.~C, 56127 Pisa, Italy}\\ 
{$^d$ Physics Department, Brookhaven National Laboratory, Upton, NY 11973, USA}
}
\begin{abstract}
On a compact space with non-trivial cycles, for sufficiently small values of the radii of the compact dimensions, SU($N$) gauge theories coupled with fermions in the fundamental representation spontaneously break charge conjugation, time reversal and parity. We show at one loop in perturbation theory that a physical signature for this phenomenon is a non-zero baryonic current wrapping around the compact directions. The persistence of this current beyond the perturbative regime is checked by lattice simulations.
\end{abstract}
\pacs{12.38.Aw, 11.30.Er, 12.38.Bx, 12.38.Gc.}
\preprint{IFUP-TH/2007-5}
\preprint{BNL-NT-07/13}
\maketitle
Quantum Chromodynamics (QCD) is the theory of strong interactions. Experimental evidence suggests that the theory is invariant under charge conjugation (C), parity (P) and time reversal (T) (see~\cite{Yao:2006px} for a recent account of experimental data). The invariance of QCD under P has been rigorously proved in~\cite{Vafa:1984xg}. One of the assumptions of the proof is Lorentz invariance, which holds in an infinite volume, but it is manifestly broken at finite temperature or in compact space, where parity can be spontaneously broken~\cite{Cohen:2001hf}. Although convincing arguments exist~\cite{Armoni:2007rf}, a proof of the invariance of QCD under T and C is still lacking.\\
Recently, it has been pointed out by the authors of~\cite{Unsal:2006pj} that C, P and T are spontaneously broken in a geometry with one compact dimension with toroidal topology for sufficiently small values of the radius of the torus when periodic boundary conditions are imposed on fermion fields. This provides a controllable mechanism for testing the consequences of the breaking of those symmetries in QCD. The order parameter is the vacuum expectation value ({\em vev}) of the Wilson loop winding in the compact direction
\beq
W = \mbox{Tr~P}e^{i \int_0^L A_\alpha \,d x^{\alpha}} \ ,
\eeq
with $\alpha$ the compact direction of size $L$, $g$ the coupling and $A_{\mu}$ the vector potential. In pure gauge SU($N$) $\langle W \rangle \propto e^{i \frac{2 \pi}{N}n}$ with $0 \le n < N$ for $L < L_{c}$ and $\langle W \rangle = 0$ for $L > L_c$, where $L_c$ is the critical value of the length of the compact direction~\cite{Kiskis:2003rd}. Modulo relabeling of the axes, the Euclidean rotated system corresponds to the theory at finite temperature and the transition that takes places at $L_c$ is the well known confinement-deconfinement phase transition~\cite{Lucini:2002ku,Lucini:2003zr}.\\
When fermions in the fundamental representation are considered, the structure of the ground state changes radically. At small radius, if the fermions have antiperiodic boundary conditions in the compact direction, $\langle W \rangle \propto 1$ (again this is the case for a system at finite temperature in the deconfined phase), while for periodic boundary conditions the Wilson loop can take two values with a non-zero imaginary part. These {\em vev} are related by complex conjugation. Each one of the two values identifies a possible vacuum of the theory. The effect of C, P and T is to interchange the vacua. Hence, in this system those symmetries are broken. For orientifold gauge theories in the large $N$ limit, which are related to QCD~\cite{Armoni:2003gp},  on a $S^3 \times S$ space as the radius of the $S$ is increased above a critical value keeping the radius of the $S^3$ small, the system regains invariance under C, P and T~\cite{Hollowood:2006cq}.\\
The arguments from which the phase structure of QCD on a finite volume is determined are based on perturbative calculations. Their validity beyond the perturbative regime has been proved by lattice simulations~\cite{DeGrand:2006qb}.\\
While the Wilson loop wrapping around the compact direction proves to be useful to characterise the phases, it is not a quantity that can be accessed directly in experiments. Physically, we expect a symmetry breaking to determine a detectable change in the properties of the system. Hence, at least one measurable quantity that is not invariant under the broken symmetries should acquire a {\em vev}. The spatial components in the compact directions of the baryonic current $j_{i} = \sum_{n=1}^{N_f} \bar{\psi}_n \mathbb{I} \otimes \gamma_{i} \psi_n$, where $\psi_{n}$ is the fermion field for flavour $n$, the sum runs over the flavour index and $\mathbb{I}$ is the identity in colour space, satisfy this requirement~\footnote{Note that the baryon density $j_0$ must always be zero, since it is invariant under $P$ and $T$. In addition, for theories with a different symmetry breaking pattern, like U($N$) or SU($4N$) with fermions in the symmetric or antisymmetric representation, the spatial part of the current is identically zero for symmetry reasons.}. Moreover, like the system, they are invariant under CP, CT and PT. This makes the $j_i$ suitable candidates as detectors of the symmetry breaking. If $\langle j_{\alpha} \rangle \ne 0$ for the compact direction $\alpha$, an observer will see a non-zero flux of baryons in that direction.\\
Using a similar ansatz to the one of~\cite{Unsal:2006pj}, we shall now show at one loop in perturbation theory that indeed the {\em vev} of the spatial current in a compact direction is different from zero. The Lagrangian for a SU($N$) gauge theory coupled with $N_f$ degenerate flavours of fermions of mass $m$ in the fundamental representation is
\beq 
\label{l0}
{\cal L} =  - \frac{1}{2 g^2} \mbox{Tr}\left( G_{\mu \nu} (x) G^{\mu \nu}(x) \right) + \sum_{n=1}^{N_f} \bar{\psi}_n (x)\left(i \dslash - \Aslash - m \right) \psi_n(x) \ , 
\eeq
where $G_{\mu \nu} = \partial_{\mu} A_{\nu} - \partial_{\nu} A_{\mu} + [A_{\mu},A_{\nu}]$. The corresponding partition function is
\beq
\label{zeta}
Z = \int \left( {\cal D} A_\mu \right) \mbox{det}\left( i \dslash - \ \Aslash - m \right)^{N_f} e^{- \frac{i}{2g^2}\int d^4 x \mbox{Tr}\left( G_{\mu \nu} G^{\mu \nu} \right)} 
\eeq
We consider the system on a $T \times L^3$ manifold, in which the $T$ direction corresponds to time and the three spatial compact directions $L$ are equal. We impose periodic boundary conditions in space, while $T$ is assumed to be large enough for the choice of boundary conditions in that direction to be irrelevant. The path integral~(\ref{zeta}) can be evaluated at one loop, by fixing a diagonal background gauge. This gives an effective one loop potential for the diagonal components of the gauge field
\beq
\label{ansatz}
\vec{A} = \left(
\begin{array}{ccc}
\frac{\vec{v}_1}{L} & & \\
& \ddots & \\
& & \frac{\vec{v}_N}{L}
\end{array}
\right) \ , \qquad \sum_{i} \vec{v}_i = 0 \ \mbox{mod}(2 \pi)  \ ,
\eeq
which reads~\cite{Barbon:2006us}
\beq
\label{effectivepot}
V(\vec{v}_1, \dots, \vec{v}_N) = \left[ \sum_{i,j = 1}^{N} f(0, \vec{v}_i-\vec{v}_j)
- 2 N_f
\sum_{i=1}^{N} f(m, \vec{v}_i) \right] \ .
\eeq
The first sum comes from the integration of the fluctuations of the gauge and ghost fields, while the second sum comes from the fermion determinant. The function $f$ is defined as
\beq
f(m,\vec{v}) =  \frac{1}{L} \left( \frac{m L}{\pi}\right)^2 \sum_{\vec{k} \neq 0} \frac{K_2( m L k)}{k^2} \sin^2 \left( \frac{1}{2} \vec{k}\cdot \vec{v} \right)  \ ,
\eeq
with the sum running over vectors in $\mathbb{Z}^3 - \vec{0}$ and $K_2$ the order two modified Bessel function of the second kind. From the asymptotic behaviour of $K_2(x)$ at small $x$, $K_2(x) = 2/x^2$, we get
\beq
f(0,\vec{v}) =  \frac{2}{\pi^2 L} \sum_{\vec{k} \neq 0} \frac{\sin^2 \left( \frac{1}{2} \vec{k}\cdot \vec{v} \right)} {k^4} \ .
\eeq
For $m \gg L^{-1}$, $K_2(m L k) \approx e^{- m L k}\sqrt{\pi/(2mLk)}$ and the sum in $f$ is dominated by terms with $k = 1$. This is true also in general, the higher frequencies in the sum being quickly oscillating with amplitude suppressed at least as $1/k^4$. With the constraints in Eq.~(\ref{ansatz}), the minima of the effective potential are located at
\beq
\label{minima}
v_1^j = v_2^j = \dots = v_N^j = \left\{
\begin{array}{l}
\pm \frac {N-1}{N} \pi \qquad \mbox{for $N$ odd}\\
\pi \qquad \qquad \ \ \mbox{for $N$ even}\\
\end{array}
\right.
\ .
\eeq
There are eight degenerate minima for odd $N$ and one minimum for even $N$. In the former case, $\langle W \rangle$ develops a {\em vev} with an imaginary part, and the spontaneous symmetry breaking occurs.\\
The baryonic current can be computed adding a source to the Lagrangian~(\ref{l0}). Defining
\beq
{\cal L}(\vec{\mu}) = {\cal L} + \vec{\mu} \cdot \vec{j} \ , 
\eeq
~\\
we obtain
\beq
\langle j_i \rangle = - i \frac{1}{L^3 T} \left( \frac{\partial \ \ }{\partial \mu_i} \log Z[\vec{\mu}] \right) _{\vec{\mu} = 0} \ ,
\eeq
with $Z[\vec{\mu}]$ the partition function in the presence of a source $\vec{\mu}$.\\
The source has the effect of shifting $\vec{v_i} \to \vec{v_i} + L\vec{\mu}$ in the expression for the effective potential~(\ref{effectivepot}). This does not change the gauge contribution. Since
\beq
Z[\vec{\mu}] = e^{i T V(\vec{v}_1 + L\vec{\mu}, \dots, \vec{v}_N + L\vec{\mu})} \ ,
\eeq
at the minima~(\ref{minima}) we get
\beq
\label{joneloop}
\langle \vec{\jmath} \rangle = - \frac{N_f N}{L^3} 
\left( \frac{m L}{\pi}\right)^2 \sum_{\vec{k} \neq 0} \frac{K_2( m L k)}{k^2} \sin \left(\vec{k}\cdot \vec{v} \right) \vec{k} \ .
\eeq
$\langle j_i \rangle$ is zero when $v_i = 0$ or  $v_i = \pi$ (i.e. when the symmetry breaking does not occur in direction $i$), is odd under $v_i \to -v_i$, goes to zero when $m \to \infty$. Hence it fulfills all the natural requirements in the current scenario. In particular, we expect a non zero current for an odd number of colours.\\  
In order to get a better handle on the properties of the baryonic current in the broken phase beyond perturbation theory, we have performed a lattice simulation using four flavours of staggered quarks coupled to an SU(3) gauge field. The number of flavours has been fixed as the minimal one for which the staggered action has an undoubtedly well defined continuum limit. For the pure gauge action we have used the standard Wilson form $ S_G = \beta \sum_{P}\left(1 - \frac{1}{3} \mbox{Tr} U_P \right) $, where $\beta = 2 N/g_0^2$ is the coupling of the theory, $U_P$ is the path-ordered product of link variables around the elementary plaquette $P$ and the sum runs over all plaquettes $P$. For the fermionic part we have used the simple staggered action
\beq
S_F = \sum_{x,\mu} \eta_{\mu}(x) \frac{1}{2} \left( \bar{\chi}(x) U_{\mu}(x) \chi(x + \hat{\mu}) - 
c.c.\right) + a m \sum_x \bar{\chi}(x) \chi(x) \ , 
\eeq
with $\eta_{\mu}(x) = (-1)^{\sum_{\nu=0}^{\mu - 1} x_{\nu}}$ ($\eta_0(x) = 1$), $\chi$ a complex three-vector, $a m$ the mass in lattice units ($a$ is the lattice spacing) and c.c. stands for the complex conjugate term to the first one in parentheses. More complicated formulations of the action or choice of another discretised form for the fermionic fields would have added extra complication with very little payback for the problem at hand.\\
\begin{figure}[h]
\includegraphics*[scale=0.35]{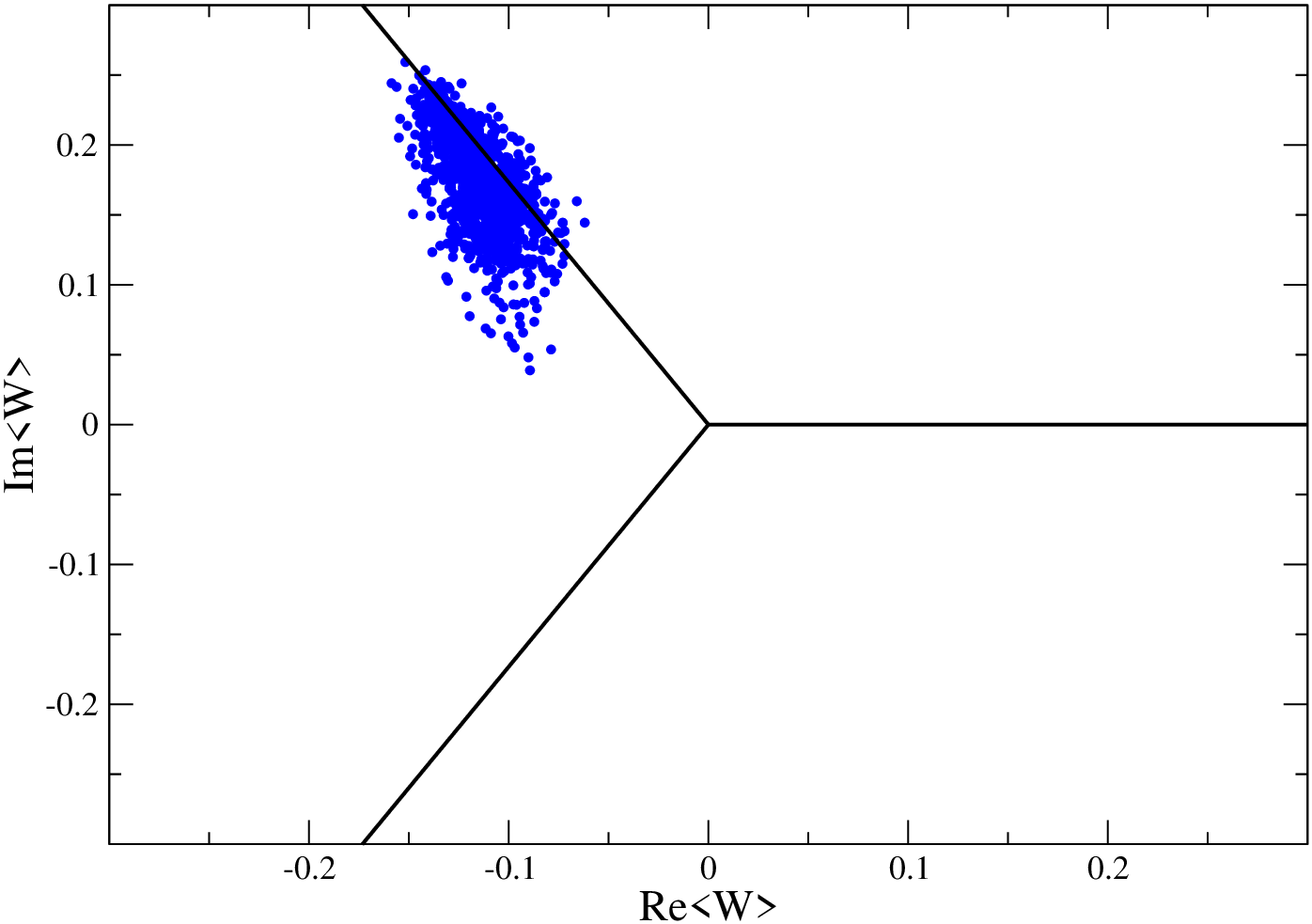}
\caption{Scatter plot for 1000 measurements of the Wilson line in one compact direction on a $24 \times 4^3$ lattice at $\beta = 5.5$ and $ a m = 0.1$. The directions corresponding to the three cubic roots of the unity are indicated by the black solid lines.}
\label{Fig:wscatter}
\end{figure}
Using as a base the publicly available MILC code~\footnote{See http://www.physics.utah.edu/${\sim}$detar/milc.}, we have performed a simulation for $\beta = 5.5$ and $a m = 0.1$. The physical scale has been determined by measuring the Sommer parameter $r_0$~\cite{Sommer:1993ce} on a $24 \times 16^3$ lattice, where the three equal spatial directions $N_s$ have been closed with periodic boundary conditions and the temporal direction $N_t$ with antiperiodic boundary conditions for the fermions, while the gauge fields are periodic in all directions. We find $a r_0 = 4.0(1)$; since the Sommer scale is $\simeq 0.5$ fm, the lattice spacing is $a \simeq 0.125$ fm, which means that $L_s = a N_s \simeq$ 2 fm and $L_t = a N_t \simeq 3$ fm. Hence, in physical units the lattice is large enough for the calculation to be reliable and the spatial volume is such that C, P and T are not broken.\\
We then studied the system with the same $\beta$ and $m$ on a $24 \times 4^3$ lattice, with the same boundary conditions as above. The spatial geometry is a three-torus, while the size of the temporal direction is large enough for the system to be confined. In this setup, $L_s \simeq 0.5$ fm. A quick check of the {\em vev} of the spatial Wilson loops shows that the system is in the broken symmetry phase. An example of the obtained distribution for $\langle W \rangle$ is displayed in Fig.~\ref{Fig:wscatter}, which shows $\langle W \rangle$ clustering around $e^{i \frac{2}{3} \pi}$.\\
\begin{figure}[h]
\includegraphics*[scale=0.35]{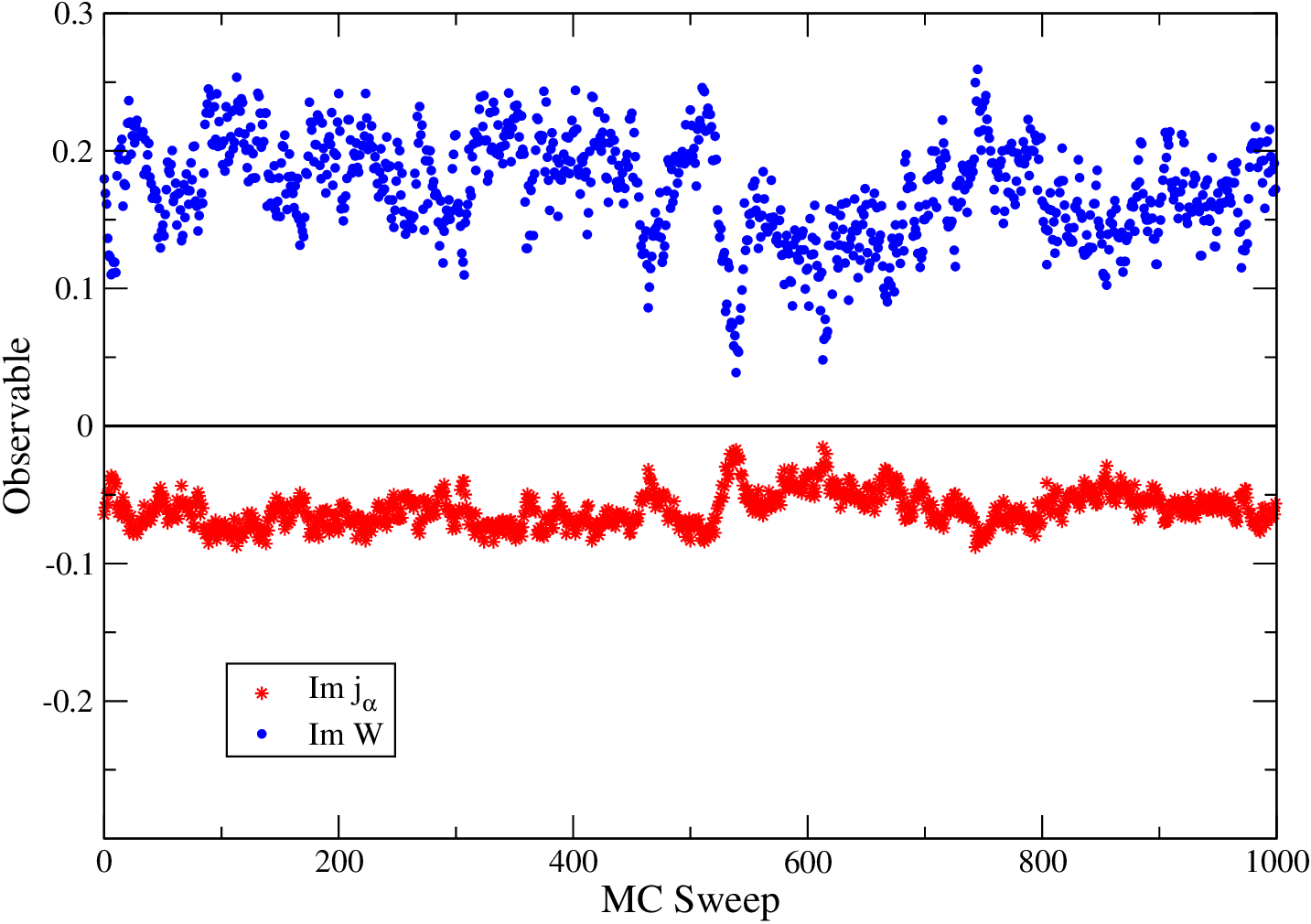}
\caption{The imaginary part of the current and the imaginary part of the Polyakov loop in one compact direction as a function of the Monte Carlo sweeps.}
\label{Fig:ksx}
\end{figure}
\begin{figure}[h]
\includegraphics*[scale=0.35]{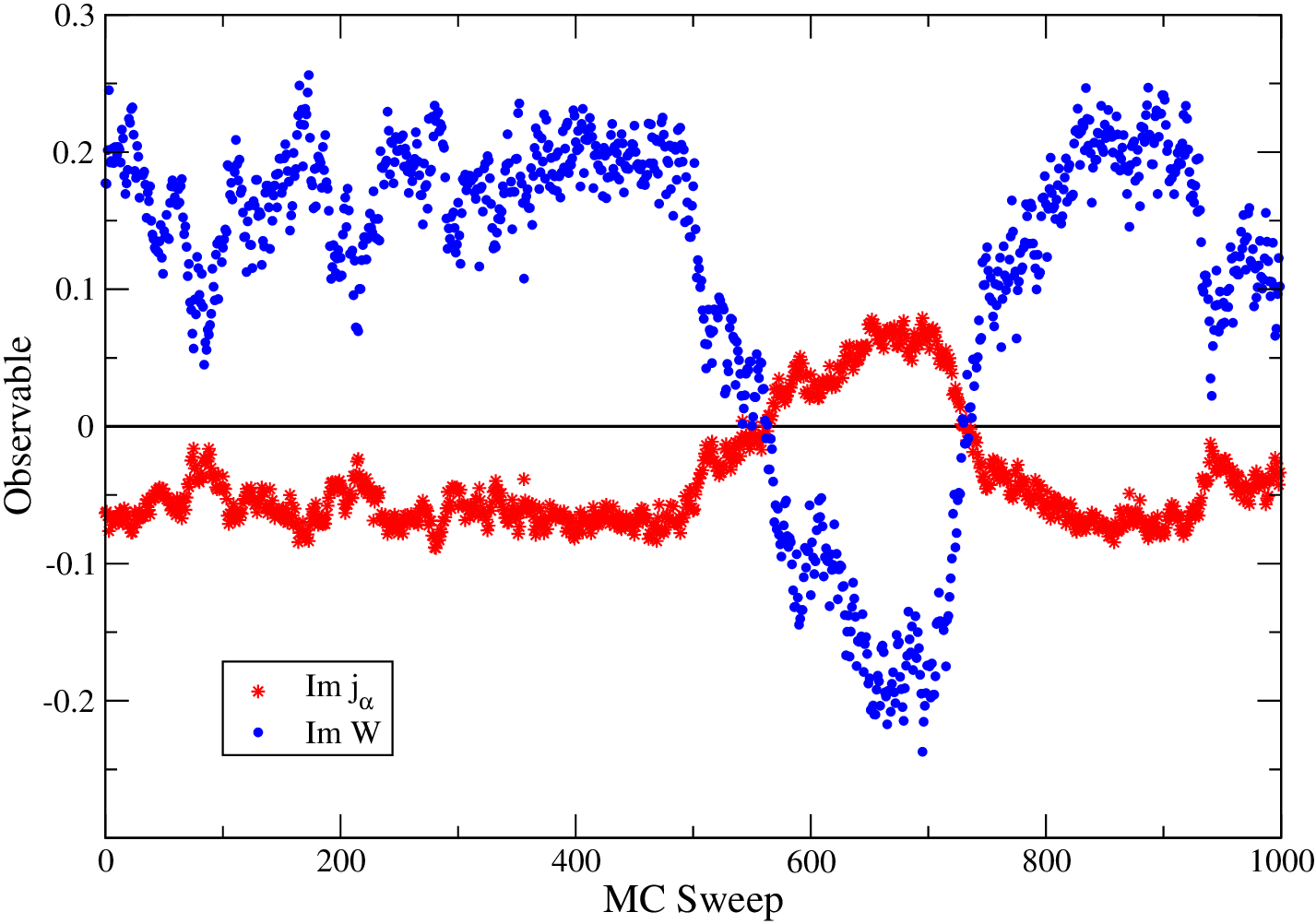}
\caption{As in Fig.~\ref{Fig:ksx}, but in another compact direction. The system shows a transition between two vacua. The transition probability is finite, due to the finite lattice extension.}
\label{Fig:ksz}
\end{figure}
The discretised version of the baryonic current can be obtained like in the continuous case, by relating the physical fermionic degrees of freedom to the staggered ones. Defining the massless Dirac operator as
\beq
D^{x,y} = \sum_{\mu} \eta_{\mu}(x) \left(U_{\mu}(x) \delta_{y,x+\hat{\mu}} - U_{\mu}^{\dag}(x-\hat{\mu}) \delta_{y,x-\hat{\mu}} \right) 
\eeq
and the four matrices
\beq
K_{\mu}^{x,y} = \eta_{\mu}(x) \left(U_{\mu}(x) \delta_{y,x+\hat{\mu}} + U_{\mu}^{\dag}(x-\hat{\mu}) \delta_{y,x-\hat{\mu}} \right)
\eeq
the current reads
\beq
\langle \vec{\jmath} \rangle = \frac{1}{TL^3} \mbox{Tr} \left((D+m)^{-1} \vec{K} \right) \ .
\eeq
In order to evaluate the current on a given configuration, we have taken 100 stochastic estimates. Since the current is an antihermitian operator in the Euclidean space, we expect its imaginary part to develop a {\em vev}, while the real part should average to zero. In Fig.~\ref{Fig:ksx} we show the behaviour of the imaginary part of the baryonic current in a compact direction as a function of the Monte Carlo sweeps, and we contrast such behaviour with that of the imaginary part of the Wilson line in the same direction. Not only does the plot show that the baryonic current is different from zero, but it also strongly suggests that there is a correlation between the value of the current and the value of the Wilson line. In particular, the modulus of the imaginary part of the current grows when the modulus of the imaginary part of the Wilson line grows, the sign being opposite between the two. This is better shown by Fig.~\ref{Fig:ksz}, which displays the behaviour of the current in another compact direction. In this case, the system makes a transition between the vacuum identified by the phase of the Wilson line being $\frac{2}{3} \pi$ to the other vacuum and then back. Noticeably, the current changes sign exactly at the points in which the imaginary part of the Wilson line changes sign, with its magnitude always tracking closely the magnitude of the phase of $\langle W \rangle$. The sum of the terms with $|\vec{k}|=1$ in the current~(\ref{joneloop}) is proportional to $\langle W \rangle$. The strong correlation between the two quantities suggests that the non-leading terms in~(\ref{joneloop}) do not affect significantly the behaviour of $j_{\alpha}$.\\
Transitions between the different vacua like those shown in Fig.~\ref{Fig:ksz} are possible because of the finite spatial size of the system.~We have verified that increasing $\beta$ at fixed lattice extensions $N_s$ and $N_t$, which corresponds to decreasing the physical volume, the frequency of the transitions increases. Likewise, decreasing $\beta$ decreases the likelihood of a transition taking place.\\
Since the baryonic current is zero for symmetry reasons in the symmetric phase, its behaviour in the broken symmetry phase makes it legitimate to use that current as an order parameter for the symmetry breaking. For consistency, we have also checked that the real part of the current in the compact directions and the zero component are zero also in the broken symmetry phase.\\
In order to evaluate the magnitude of the current, we averaged over directions for which no tunneling between the two vacua took place. We find
\beq
|\mbox{Im}\langle j_{\alpha} \rangle| = 0.060 \pm 0.002 \ .
\eeq
It is instructive to compare this number with the one loop expression, Eq.~(\ref{joneloop}), which gives $\langle j_{\alpha} \rangle \simeq 0.037473(4)$, where the error is a conservative estimate for the truncation of the sum. Hence, quite remarkably the one loop calculation pins down the correct order of magnitude even for a compact dimension with size of the order of $1/\Lambda_{QCD}$. Non-perturbative effects could explain the discrepancy between the perturbative formula and the measured value. Besides, our calculation being at one single lattice spacing, we do not have any handle on the size of discretisation errors. For this reason, a careful comparison between the perturbative expression and the lattice result should be the subject of a more detailed study, which is beyond the scope of this paper. Our preliminary Monte Carlo results for the current closer to the continuum limit show substantial agreement between the measured value and the perturbative formula.\\
In conclusion, we have shown that QCD on small compact dimensions with non-trivial cycles is characterised by a flow of current whose sign depends on the vacuum selected by the system. The persistent baryonic current reminds the supercurrent observed in superconductors. However, there is a fundamental difference: unlike the case of superconductors, in QCD in compact not simply connected space the current is still conserved, since the U(1) baryon symmetry (which in the case of QCD is a global symmetry) remains unbroken. The persistent flow is induced by the spontaneous breaking of a discrete symmetry, charge conjugation. The baryonic current can be used as an order parameter for the spontaneous breaking of charge conjugation in SU($N$) gauge theories. Over the Wilson line, it has the advantage of being an observable quantity. Moreover, unlike the Wilson line, which is ultraviolet divergent on the lattice, the baryonic current is a well defined observable. This makes it better suited for numerical studies of the physics of C parity spontaneous breaking close to the continuum limit. A similar investigation is currently in progress.\\
~\\
{\bf Acknowledgments:}
We would like to thank A. Armoni, T. Hollowood and C. Hoyos for useful insight on their works. We also thank G. Aarts, L. Del Debbio, A. Di Giacomo, S. Hands and F. Karsch for interesting discussions, and M. Unsal and L. Yaffe for useful comments. The numerical simulations were performed on a 15 dual-core dual-processor AMD Opteron Cluster partially funded by The Royal Society and PPARC at the University of Swansea. The work of B.L. has been supported by the Royal Society. The work of C.P. has been supported in part by contract DE-AC02-98CH1-886 with the U.S. Department of Energy.

\bibliographystyle{apsrev}
\bibliography{cqcd}
\end{document}